# Defect in Photonic Time Crystals

Snehashis Sadhukhan [1] and Somnath Ghosh [1,*]
*Department of Physics, Indian Institute of Technology Jodhpur, Rajasthan- 342030, India*
*Corresponding author: somiit@rediffmail.com*

Photonic Time Crystals (PTCs) provide a completely new platform exhibiting light wave amplification owing to periodically varying electromagnetic properties. The need to control this amplification is becoming increasingly important, especially with the emergence of meta surface based practical realization of PTCs. The work introduces isolated temporal defect in PTCs to establish a new degree of control over the amplification. We find that in presence of the defect, the transmittance and reflectance become close to unity for a specific value of momentum ($k_d$) within the bandgaps accompanied by a significant impact on the amount of amplification. We show the impact of the temporal defect on the exponential growth of intensity with PTC periods. The effect primarily depends on the Floquet frequency of the PTC that becomes real at $k_d$ giving rise to four pulses instead of two as an outcome of gap propagation. We further demonstrate that by manipulating the temporal and dielectric properties of the defect, the defect state in momentum can be tuned to serve the design interest for specialty applications.

In the recent past, due to extraordinary advancements in theoretical and experimental studies, photonic time crystals (PTC) have emerged as a new platform for electromagnetic wave amplification in photonics [1–7]. Owing to the periodic sharp temporal change of refractive index ($n$) [8], PTCs produce momentum ($k$) gaps in the dispersion diagram that give rise to the exponential growth in amplitude of a propagating pulse [4,9]. The concept of $k$ gaps in a temporally varying system has already been demonstrated in the radio frequencies (RF) [10]. In such time varying systems, the propagating wave draws energy from the pumping wave that is used to modulate the PTC. Pumping waves have been demonstrated to transmit energy to a travelling wave in microwave domain while passing through a temporally modulated transmission line [6,11,12]. Recently, the exponential growth of electromagnetic waves has been demonstrated in time varying meta surface for microwave domain [3]. Owing to such exceptional advancements, the unconventional non-resonant exponential amplification in PTCs has paved the way for PTC lasers and amplifiers [1,2].

When a propagating wave encounters a temporal boundary in PTC, it undergoes time reflection and generate a transmitted wave along with its time-reflected pair while conserving their momentum [4,9,13]. The strong interaction among these periodically generated transmitted and time-reflected waves results in several novel characteristic features that include momentum gaps [4,8,9,14,15], topological features [9,16], pulse localization in disorder [13], reciprocity at temporal edges [2], momentum dependent amplification [4] etc. An optical pulse propagating through a PTC with $k$ lying within the $k$ gap gets localized and amplified. On the other hand, such effects are absent for $k$ of the pulse lying within a band. In the $k$ gap the Floquet frequencies ($\Omega$) becomes complex and the envelope function of the modes with $\pm\Omega$ manifests exponential amplifications. Such amplification in PTC is found to be dependent on the value of $k$ in the gaps as well. Moreover, the maximum amount of amplification varies bandgap to bandgap [4]. Thus, tuning the amplification in PTCs is of immense interest to proceed further for amplification or lasing applications.

For a long time, the optical regime of time varying media was unexplored and very few theoretical investigations have been reported. However, starting from 2016 [8], the interest in PTCs has grown rapidly and realizing the potential opportunities PTC will bring on the table for all optical devices, and now attempts are being made to realize PTCs. Observing such unique properties in optical regime requires a change in $n\sim1$ in a time scale of fs that generates a significant $k$ gap [9,17]. To realize such fast and abrupt index changes, experimentalist are looking into promising materials such as TCOs, MoSe$_2$, and transdimensional films [18,19]. Moreover, efforts are being made to reduce the modulation power while realizing the PTCs. Thus, it is only a matter of time, when PTCs will be demonstrated as non-resonant amplifiers in the optical regime, and it is the need of the time to emphasis on the tunability of such amplification [20].

Here we study the control over the amplification in PTCs. We introduce isolated temporal defect in PTCs and study the pulse dynamics in presence of the defect. We find that the amplification becomes close to unity for a single value of $k$ within the $k$ gap. Finally, we engineer the defect to explore as an additional degree of freedom to emphasize control over the amplification in PTCs.

We consider a plane wave propagating through a spatially homogeneous medium where $n$ varies periodically with time as shown in Fig. 1(a) and begin with developing a generalized analytical formalism. Such binary and instantaneous variation depicted in the schematic representation leads to fundamental physical effects in connection to the defect in PTCs, without the loss of generality [9,21]. The band diagram in general, under such circumstances can be obtained using the transfer matrix method as discussed in [4]. The displacement field takes the form written as $D_x(z,t) = a_p e^{i\omega_1(t-pT)} + b_p e^{-i\omega_1(t-pT)}$ for $pT - t_1 < t < pT$, or $= c_p e^{i\omega_2(t-pT+t_1)} + d_p e^{-i\omega_1(t-pT)}$ for $pT - t_1 < t < pT$ and the dispersion relation describing the Floquet frequency ($\Omega$) as a function of $k$ for infinite number of temporal bi-layers turns out to

be
$$\Omega(k) = \frac{1}{T} \cos^{-1} \frac{P+Q}{2} \quad (1)$$

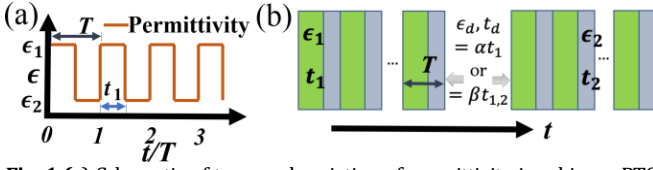

**Fig. 1** (a) Schematic of temporal variation of permittivity in a binary PTC showing step like variation from $\epsilon_1 = 3$ to $\epsilon_2 = 1$ and vice versa with time periodicity of $T$. (b) Schematic of an isolated temporal defect sandwiched in between two identical PTCs.

where, $T$ is the time period of the PTC, $P = e^{i\omega_2 t_2}(\cos(\omega_1 t_1) + \frac{i}{2}\left(\frac{\omega_1}{\omega_2} + \frac{\omega_2}{\omega_1}\right)\sin(\omega_1 t_1))$, $Q = \frac{i}{2} e^{i\omega_2 t_2}\left(\frac{\omega_1}{\omega_2} + \frac{\omega_2}{\omega_1}\right)\sin(\omega_1 t_1)$ and $\omega_{1,2} = \frac{k}{n_{1,2}}$. Truncating the number of temporal bi-layers up to $N$ leads to modification in the matrix equations and the same formalism can be opted to obtain the transmittance or reflectance which quantifies the amplification in PTCs with $N$ number of temporal unit cells. To get hold of the amplification, we introduce an isolated temporal defect sandwiched in between two identical PTCs with $N$ temporal unit cells. For simplicity, we consider the defect with permittivity $\epsilon_d$ and of temporal width $t_d$ [see Fig. 1(b)]. A plane wave propagating through such a system, will face a PTC for a duration of $NT$ and then undergo a free space propagation for $t_d$ before entering the next PTC. As a result, the transfer matrix in this case will be modified as,

$$\begin{pmatrix} c_0 \\ d_0 \end{pmatrix} = \begin{pmatrix} P & Q \\ R & S \end{pmatrix}^N \begin{pmatrix} e^{-i\omega_d t_d} & 0 \\ 0 & e^{i\omega_d t_d} \end{pmatrix} \begin{pmatrix} P & Q \\ R & S \end{pmatrix}^N \begin{pmatrix} c_{2N} \\ d_{2N} \end{pmatrix} \quad (2)$$

where, $S = P^*$, $R = Q^*$ and $\omega_d = \frac{k}{n_d}$, $n_d$ is the refractive index of the defect. The defect matrix signifies the propagation through the homogeneous defect, where $e^{\pm i\omega_d t_d}$ stands for wave propagation in the defect where there is no time-reflection. To study the effect of the defect on overall band structure and amplification, we consider $\epsilon_1 = 3$ and $t_1 = 1$ fs, $\epsilon_2 = 1$ and $t_2 = 1$ fs respectively resulting in a time period of $T = 2$ fs. At first, we consider a PTC with $N = 10$ and plot of $\text{Im}(\Omega)$ vs $k$ as shown in Fig. 2(a). Now we truncate the PTCs at $N = 5$ each, and generate the plot of $\text{Im}(\Omega)$ vs $k$ [shown in Fig. 2(b)] which reveals that due to presence of the defect within the bandgap we get a value of $k$ ($=k_d$) where the $\Omega$ becomes completely real within bandgap. Within every bandgap such a $k_d$ value is present, which in case of the given PTC system turns out to be, $k_d = 0.60$ in the first bandgap and $k_d = 1.30$ in the second bandgap. In absence of a defect, $\Omega$ is real within the band where we obtain no amplification and complex at the bandgaps resulting in exponential growth in wave amplitude. Thus, it is evident that the defect has introduced an additional state within each of the bandgaps where the amplification should be close to unity.

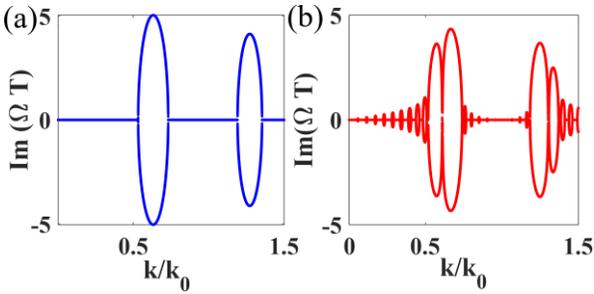

**Fig. 2** Imaginary part of Floquet frequency ($\Omega$) plot with normalized momentum ($k_0 = \frac{2\pi}{Tc}$, $c$ is velocity of light in free space) for (a) PTC without defect showing $\Omega$ is complex only within the bandgap, (b) PTC with defect showing real value of $\Omega$ within the bandgaps.

To observe the effect on amplification in presence of defect, within the bandgap, the transmittance and reflectance plots are obtained and compared with the same in absence of the defect. The transmittance and reflectance in case of PTCs quantifies the amount of amplification the transmitted and time reflected waves, respectively [4]. In presence of defect, the transmittance which is obtained from the transmission coefficient gets modified since,

$$\begin{pmatrix} c_0 \\ 0 \end{pmatrix} = \begin{pmatrix} P & Q \\ R & S \end{pmatrix}^N \begin{pmatrix} e^{-i\omega_d t_d} & 0 \\ 0 & e^{i\omega_d t_d} \end{pmatrix} \begin{pmatrix} P & Q \\ R & S \end{pmatrix}^N \begin{pmatrix} c_{2N} \\ d_{2N} \end{pmatrix} = \begin{pmatrix} W & X \\ Y & Z \end{pmatrix}\begin{pmatrix} c_{2N} \\ d_{2N} \end{pmatrix} \quad (3)$$

and the transmittance and reflectance are

$$|t_N|^2 = \frac{1}{\left|W - \frac{XY}{Z}\right|^2}, |r_N|^2 = \frac{1}{\left|X - \frac{WZ}{Y}\right|^2} \quad (4)$$

For the PTC under consideration, the transmittance plots are obtained for both cases: in absence of defect [as shown in Fig. 3(a)] and with the defect [as shown in Fig. 3(b)]. On comparing Figs. 3(a) and (b), it is evident that a propagating wave in presence of defect in the PTC has no influence of the defect in transmittance within the band. Whereas a wave with momentum within the bandgap (for example, first bandgap for $0.514 < \frac{k}{k_0} < 0.731$ and second bandgap $1.191 < \frac{k}{k_0} < 1.353$), in absence of defect, suffers weaker amplification at the edges and amplification increases as we move towards the center of the gap. Interestingly, in presence of defect, there exists a drastic fall in amplification around the $k_d$ (=0.60 $k_0$ and 1.30 $k_0$ for the first and second bandgap, respectively) within a small periphery in the bandgap and the transmittance reaches a value close to unity at $k_d$. In both sides of $k_d$, we observe an increasing value of transmittance, which reaches a maximum and again reaches value close to unity at both edges of the bandgap. Moreover, the defect also reduces the maximum amount of amplification that can be achieved within a respective bandgap almost by an significant amount. As in case of our example, at the first bandgap, in absence of defect, for any value of momentum the wave is amplified. But in presence of defect, for momentum value ($k_d$) 0.60 $k_0$ we get no amplification of the propagating wave. Additionally, the presence of defect reduces the maximum amplification within the first bandgap from $\sim 6 \times 10^3$ to $\sim 1.5 \times 10^3$. The reflectance plot also reveals the same scenario, where for $k_d = 0.60 k_0$ within the first bandgap there is no amplification for the time reflected wave as well [as can be seen from Fig. 3(c) and (d)].

To unveil the exclusive features under lying as an influence of the defect we observe the pulse dynamics of the PTC where we implement finite difference time domain (FDTD) method. Here, by following Yee's grid we discretize the space and time numerically and solve Maxwell's equation to observe the evolution of a propagating pulse in time.

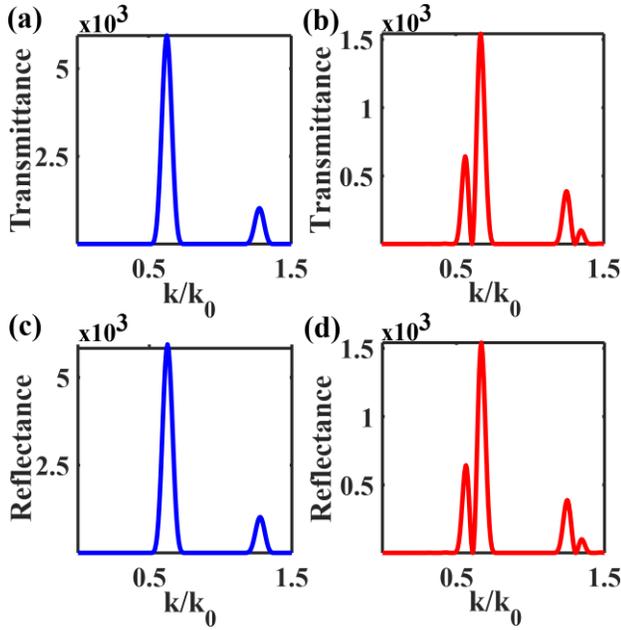

**Fig. 3** (a, b) Transmittance plot showing amplification of the time-reflected wave for (a) PTC without defect, (b) PTC with defect with normalized momentum revealing no amplification for a single value of momentum within the bandgaps where the Ω becomes real. (c, d) Reflectance plot showing amplification of the time-reflected wave for (c) PTC without defect, (d) PTC with defect with normalized momentum revealing no amplification for a single value of momentum within the bandgaps where the Ω becomes real.

To further analyze the effect of defect on PTC, we must investigate the characteristics and behavior of light inside the defect within the momentum bandgap. We simulate the propagation of three pulses inside the PTC. All of them are with full width half maxima of 50 fs. While the first pulse has a wavelength of ~1.10 $\mu$m that falls at the left side of the $k_d$ [shown in Fig. 4(a)], whereas the second pulse has a wavelength of ~ 1.00 $\mu$m [shown in Fig. 4(b)] and the third pulse has wavelength of ~ 0.92 $\mu$m [shown in Fig. 4(c)] which falls at the right side of $k_d$ at the first bandgap. At first, the pulse propagates through the free space ($\epsilon$=1), at time $t$=200 fs, then a PTC starts with the parameters described above and completes five cycles (or temporal unit cells). Here, the defect comes into play for a time $t_d$ (= 1 fs) with permittivity $\epsilon_d$ (= 1) and again the pulse enters an identical PTC at $t$ = 211 fs and ultimately leaves the PTC after it completing five cycles at $t$ = 221 fs.

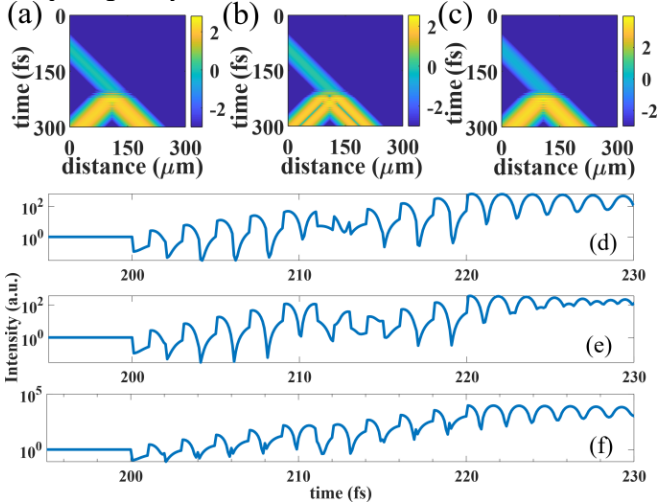

**Fig. 4** FDTD simulation of pulses propagating through a PTC with defect showing intensity variation for the momentum ($k/k_0$) value being (a) 0.55 (left side of $k_d$), (b) 0.60 (=$k_d$), (c) 0.65 (right side of $k_d$). The pulse in (b) splits into four pulses with a lower amount of amplification. Whereas, in (a), (c) pulses undergo higher amount of amplifications and splitting into two pulses as if the defect doesn't exist. (d, e, f) FDTD simulation of pulses propagating through a PTC with defect showing amplitude of displacement field with time for the momentum ($k/k_0$) values being (d) 0.55 (left side of $k_d$), (e) 0.60 (=$k_d$), (f) 0.65 (right side of $k_d$) within the first bandgap. The amplification with each temporal period stops for a while when the defect appears in case of (e) and remains almost equal for upcoming few periods. Whereas, in (d), (f) pulses starts getting amplified immediately after the defect is over.

The pulse propagating with a momentum within the bandgap but having $\frac{k}{k_0}$ value lying right to $k_d$ produces two propagating pulses at the end of the PTC, one time reflected and other is transmitted, where both are equally amplified [Fig. 4(c)]. A similar scenario is observed when the pulse has $\frac{k}{k_0}$ ling left to the $k_d$, it also produces two pulses at the end of the PTC one transmitted and one time reflected [shown in Fig. 4(a)]. In these two cases, the propagation turns out to be as if there was no defect in the PTC. Whereas, the scenario is different if the pulse travels with a momentum $k_d = 0.60\ k_0$. In this case the impact of defect in the pulse propagation is clearly evident, where we observe four pulses, two transmitted and two time reflected. This is due to the fact that in presence of defect, and at the defect location in time, Ω becomes real. As a result, at that point the pulse splits into two pulses one forward propagating and one time reflected. After the defect in time is over, these two pulses enter the second PTC and produces their own transmitted and time reflected counterpart.

For the intriguing physics, the intensity plot of the propagating pulse is observed with time. As we move towards the momentum value $k_d$ from either side in terms of momentum, we observe less growth in intensity of the pulse. The intensity variation with time for different values of $k$ around the $k_d$ brings out the effect of defect on amplification of the pulse with enhanced clarity. The plots clearly establish that when the pulse enters the PTC, with each unit cell it passes, it gets amplified. Here, the energy from the modulation is getting transferred to the propagating pulse as we observe from Fig. 4. For our PTC system, the defect appears after 5 temporal unit cell i.e., $N = 5$ from $t = 200$ fs to $t = 210$ fs, and till then the pulse gets amplified. At $t = 210$ fs when the defect appears, the pulse amplification is on hold till the next PTC starts at $t = 211$ fs. In case of pluses with momentum $0.55k_0$ [in Fig. 4(d)] and $0.65k_0$ [in Fig. 4(f)] start getting amplified in accordance the their corresponding Ω, from the beginning of the next unit cell in time. But if the pulse propagates with the momentum of 0.60 (= $k_d$) [as shown in Fig. 4(e)], the amplification is on hold for a longer time span and it is not amplified for next 2-to-3-unit cells of the second PTC. Moreover, it can be observed from Fig. 4(d) and (f) that in presence of the defect, the pulse does not feel any influence of amplification until or unless the value of momentum is $k_d$ [shown in Fig. 4(e)].. Only for this specific value of momentum, the pulse sees no change or fluctuations in intensity for the next bilayer in time. In all other cases, the Ω being complex, fluctuations in pulse intensity is present. In other words, the PTC systems in presence of a defect, stops pumping energy to the pulse only if the momentum is equals $k_d$. In this case, the propagating pulse sees the time varying system as two separate PTCs, one before and another after the presence of defect where its Ω becomes real.

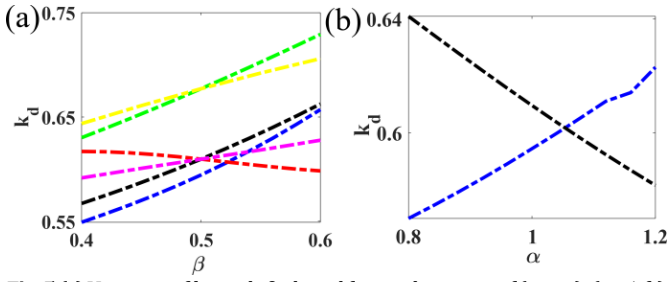

**Fig. 5** (a) Variation of $k_d$ with $\beta$ plotted for combinations of $(\epsilon_d, t_d)$: $(\epsilon_2, 1\text{ fs})$ in pink, $(\epsilon_1, 1\text{ fs})$ in yellow, $(\epsilon_1, t_1)$ in blue, $(\epsilon_1, t_2)$ in green, $(\epsilon_2, t_1)$ in red, $(\epsilon_2, t_2)$ in black. (b) Variation of $k_d$ with $\alpha$ plotted for combinations of $(\epsilon_d, t_d)$: $(\epsilon_1, \alpha t_1)$ in blue and $(\epsilon_2, \alpha t_1)$ in black.

We call attention to explore the exclusive features of the defects in the characteristic dynamics of the PTCs. A defect in PTC, in principle, could be designed of any temporal size and could be chosen to have any wide variety of permittivity. As a result, it is important to study the dependence of $k_d$ on 1D temporal defect as well as PTC parameters [as shown in Fig. 7]. We first define $\beta$ as, $t_1 = \beta T$ and $t_2 = (1 - \beta)T$ which directly tunes the PTC parameters i.e., the temporal widths of high and low permittivity regions. For a given $t_d$ (=1 fs), the dielectric property $\epsilon_d$ can take high or low values, such as $\epsilon_1$ or $\epsilon_2$. Thus, by varying $\beta$ in the range of $[0.4, 0.6]$, the variation in $k_d$ for $\epsilon_d = \epsilon_2$ at the first bandgap has been plotted in Fig. 5(a) (pink lines). A similar study has been carried out with $\epsilon_d = \epsilon_1$ in Fig. 5(a) (yellow lines). In both the cases, the $k_d$ moves to higher values of momentum, but with a higher value of slope for a chosen higher value of dielectric constant of the defect. However, when $t_d$ itself considered as a function of $\beta$ i.e., when we define $t_d$ as $= \alpha t_{1,2}$, (where $\alpha$ is the temporal width parameter of the defect) giving rise to four different cases to be observed, those for the same range of $\beta$ are plotted for $\alpha = 1$ in Fig. 5(a) as, $\epsilon_d = \epsilon_1$ and $t_d = t_1$ in blue lines, $\epsilon_d = \epsilon_1$ and $t_d = t_2$ using green lines, $\epsilon_d = \epsilon_2$ and $t_d = \alpha t_1$ in red lines and $\epsilon_d = \epsilon_2$ and $t_d = t_2$ in black lines respectively. For their combinations of $(\epsilon_d, t_d)$ i.e., $(\epsilon_1, t_1)$, $(\epsilon_1, t_2)$ and $(\epsilon_2, t_2)$ the $k_d$ is found to shift to higher values as $\beta$ moves from 0.4 to 0.6. The trace of $k_d$ is accompanied by a greater slope if the permittivity of the defect is high and becomes maximum if $t_d$ is a function of $t_1$ as well. The most interesting case appears when $\epsilon_d = \epsilon_2$ and $t_d = t_1$. In this case, the $k_d$ is found to shift to lower values of momentum as $\beta$ moves from 0.4 to 0.6. Further, the variation in $\alpha$ introduces the turnability of the temporal width of defect without changing the permittivity $\epsilon_d$. Keeping parameters of the PTCs constant, by varying $\alpha$ in the range [0.8, 1.2] the $k_d$ can be tuned further, and the respective plot is shown in Fig. 5(b) for the first bandgap. The figure shows the variation of $k_d$ for high value of permittivity of the defect ($\epsilon_d = \epsilon_1$) in blue and for low value of permittivity ($\epsilon_d = \epsilon_2$) in black. It is evident from the plot that for $\epsilon_d = \epsilon_1$, the $k_d$ can be shifted towards higher values if the value of $\alpha$ is increased. However, the increment in $\alpha$ results in lower values of $k_d$ for $\epsilon_d = \epsilon_2$. Thus, defect in the PTCs can be tuned to achieve $k_d$ at any momentum that would be dictated by the nature of defect. Further we extend our study to the effect of multiple isolated temporal defects in PTCs. The introduction of two identical isolated defects among three identical PTCs, do not introduce any additional defect state ($k_d$) within the bandgaps. As a result, the same scenario as it was obtained in case of single defect, gets repeated. The pulse, when moves with a momentum $k_d$, first gets amplified within the first PTC. Then when it encounters the defect, the amplification comes to a halt and the $\Omega$ being real at that moment, generates two pulses. These two pulses, again gets amplified within the second PTC until they face the second defect. On encountering the second defect, four pulses are generated and these four pulses after propagating through the third PTC, result in eight pulses, four time reflected and four transmitted and so on.

Thus, the temporal defects in 1D PTCs can be introduced as a very important and attractive tool to control electromagnetic wave amplification. A PTC gives rise to a range of momenta for which light is amplified while passing through the crystal. A defect by leading to a real value of $\Omega$ for a single momentum within bandgaps produce close to unity amplification in the reflection/transmission plot. A temporal defect, in PTCs, can be designed to be of any temporal size and can host a variety of design characteristics. As a result, defects within the momentum gaps can be tuned to any momentum and temporal extent of design interest. With the demonstration of momentum gaps and their amplification property of 2D PTCs in microwave domain [3], a new dimension in wireless communications facilitating amplified robust signal transmissions has been introduced. Thus, the proposed defect induced control over the amplification will add a new degree of freedom in several applications those include manipulation of light in integrated photonic circuits, wireless communications etc.

**Funding.** SS acknowledges support by Prime Minister Research Fellows (PMRF) Scheme (PMRF ID: 2201322), India.

References

[1] M. Lyubarov, Y. Lumer, A. Dikopoltsev, E. Lustig, Y. Sharabi, and M. Segev, "Amplified emission and lasing in photonic time crystals," Science 377, 425–428 (2022).

[2] A. Dikopoltsev, Y. Sharabi, M. Lyubarov, Y. Lumer, S. Tsesses, E. Lustig, I. Kaminer, and M. Segev, "Light emission by free electrons in photonic time-crystals," Proc. Natl. Acad. Sci. 119, e2119705119 (2022).

[3] X. Wang, M. S. Mirmoosa, V. S. Asadchy, C. Rockstuhl, S. Fan, and S. A. Tretyakov, "Metasurface-based realization of photonic time crystals," Sci. Adv. 9, eadg7541 (2023).

[4] S. Sadhukhan and S. Ghosh, "Momentum controlled optical pulse amplification in photonic time crystals," arXiv:2301.09390 (2023).

[5] E. Galiffi, R. Tirole, S. Yin, H. Li, S. Vezzoli, P. A. Huidobro, M. G. Silveirinha, R. Sapienza, A. Alù, and J. B. Pendry, "Photonics of time-varying media," Adv. Photonics 4, (2022).

[6] J. G. Gaxiola-Luna and P. Halevi, "Growing fields in a temporal photonic (time) crystal with a square profile of the permittivity ε(t)," Appl. Phys. Lett. 122, 011702 (2023).

[7] Z. Hayran, J. B. Khurgin, and F. Monticone, "ℏω versus ℏk: dispersion and energy constraints on time-varying photonic materials and time crystals [Invited]," Opt. Mater. Express 12, 3904 (2022).

[8] A. M. Shaltout, J. Fang, A. V. Kildishev, and V. M. Shalaev, "Photonic Time-Crystals and Momentum Band-Gaps," in Conference on Lasers and Electro-Optics (OSA, 2016), p. FM1D.4.

[9] E. Lustig, Y. Sharabi, and M. Segev, "Topological aspects of photonic time crystals," Optica 5, 1390 (2018).

[10] J. R. Reyes-Ayona and P. Halevi, "Observation of genuine wave vector ( k or β ) gap in a dynamic transmission line and temporal photonic crystals," Appl. Phys. Lett. 107, 074101 (2015).

[11] P. K. Tien, "Parametric Amplification and Frequency Mixing in Propagating Circuits," J. Appl. Phys. 29, 1347–1357 (1958).
[12] E. S. Cassedy, "Dispersion relations in time-space periodic media part II—Unstable interactions," Proc. IEEE 55, 1154–1168 (1967).
[13] Y. Sharabi, E. Lustig, and M. Segev, "Disordered Photonic Time-Crystals," Phy. Rev. Let. 126, 163902 (2021).
[14] E. Lustig, O. Segal, S. Saha, C. Fruhling, V. M. Shalaev, A. Boltasseva, and M. Segev, "Photonic time-crystals - fundamental concepts [Invited]," Opt. Express 31, 9165 (2023).
[15] J. T. Mendonça and P. K. Shukla, "Time Refraction and Time Reflection: Two Basic Concepts," Phys. Scr. 65, 160–163 (2002).
[16] J. Ma and Z.-G. Wang, "Band structure and topological phase transition of photonic time crystals," Opt. Express 27, 12914 (2019).
[17] E. Lustig, S. Saha, E. Bordo, C. DeVault, S. N. Chowdhury, Y. Sharabi, A. Boltasseva, O. Cohen, V. M. Shalaev, and M. Segev, "Towards photonic time-crystals: observation of a femtosecond time-boundary in the refractive index," in Conference on Lasers and Electro-Optics (Optica Publishing Group, 2021), p. FF2H.1.
[18] S. Saha, O. Segal, C. Fruhling, E. Lustig, M. Segev, and A. Boltasseva, "Photonic time crystals: A materials perspective," Opt. Express 31, 8267-8273 (2023).
[19] W. Jaffray, S. Saha, V. M. Shalaev, A. Boltasseva, and M. Ferrera, "Transparent conducting oxides: from all-dielectric plasmonics to a new paradigm in integrated photonics," Adv. Opt. Photonics 14, 148 (2022).
[20] S. Sadhukhan and S. Ghosh, "Bandgap engineering to control amplification in photonic time crystals," in Frontiers in Optics + Laser Science 2022 (FIO, LS) (Optica Publishing Group, 2022), p. JTu5B.9.
[21] F. Biancalana, A. Amann, A. V. Uskov, and E. P. O'Reilly, "Dynamics of light propagation in spatiotemporal dielectric structures," Phys. Rev. E 75, 046607 (2007).